\documentclass[aps,onecolumn,10pt,superscriptaddress,floatfix,nofootinbib]{revtex4}
\usepackage{subfigure}
\usepackage{epsfig}
\usepackage{epstopdf}
\usepackage{color}
\usepackage{graphicx}
\usepackage{longtable}
\usepackage{CJK}
\usepackage{amsmath,bbold}
\usepackage{float}
\usepackage{physics}
\usepackage{comment}

\usepackage{hyperref}
\hypersetup{colorlinks=true,linkcolor=red, citecolor=blue,  urlcolor=blue}
\makeatletter
\def\@yfootnote{\stepcounter\@mpfn\protected@xdef\@thefnmark{\thempfn}\@footnotemark}
\makeatother

\usepackage{listings}
\usepackage{xcolor}

\definecolor{codegreen}{rgb}{0,0.6,0}
\definecolor{codegray}{rgb}{0.5,0.5,0.5}
\definecolor{codepurple}{rgb}{0.58,0,0.82}
\definecolor{backcolour}{rgb}{0.95,0.95,0.92}

\lstdefinestyle{mystyle}{
    backgroundcolor=\color{backcolour},
    commentstyle=\color{codegreen},
    keywordstyle=\color{magenta},
    numberstyle=\tiny\color{codegray},
    stringstyle=\color{codepurple},
    basicstyle=\ttfamily\footnotesize,
    breakatwhitespace=false,
    breaklines=true,
    captionpos=b,
    keepspaces=true,
    numbers=left,
    numbersep=5pt,
    showspaces=false,
    showstringspaces=false,
    showtabs=false,
    tabsize=2
}

\begin{document}

\title{\texttt{u4lib}: An AI-Ready Computational Framework for U(4) Symmetry Algebra,\\ Wigner Coupling, and Racah Recoupling Coefficients}

\author{Phong Dang}
\affiliation{Department of Physics and Astronomy, Louisiana State University, Baton Rouge, LA 70803, USA}

\author{Daniel Langr}
\affiliation{Department of Computer Systems, Faculty of Information Technology, Czech Technical University in Prague, Prague 16000, Czech Republic}

\author{Tomas Dytrych}
\affiliation{Nuclear Physics Institute, Czech Academy of Sciences, 250 68 \v{R}e\v{z}, Czech Republic}
\affiliation{Department of Physics and Astronomy, Louisiana State University, Baton Rouge, LA 70803, USA}

\author{Michela Negro}
\affiliation{Department of Physics and Astronomy, Louisiana State University, Baton Rouge, LA 70803, USA}

\begin{abstract}
Wigner's supermultiplet symmetry treats nucleon spin and isospin within a
common U(4)/SU(4) description. The C++ library \texttt{u4lib} supplies the
coupling coefficients and algebraic operations needed for calculations in this
scheme. 
Users can construct irreducible representations (irreps), obtain dimensions and Casimir
invariants, conjugate states, enumerate the canonical Gelfand-Tsetlin and physical 
$\mathrm{SU_S(2)}\otimes\mathrm{SU_T(2)}$ bases with inner multiplicities, and decompose
tensor products. Matrix-element routines cover the canonical U(4) generators
and the physical spin $\mathbf{S}$, isospin $\mathbf{T}$, and
Gamow-Teller-type tensor $\mathbf{U}$. Wigner coupling and Racah recoupling
coefficients are evaluated by the 2024 procedure of Dang \textit{et al.} 
The implementation builds on \texttt{u3lib}, with a
header-only C++ interface and corresponding Python bindings. A catalog of
agent skills assists an AI agent in preparing, executing, and checking
calculations through natural-language requests. 

\end{abstract}


\maketitle

\section*{PROGRAM SUMMARY}
\noindent
\textit{Program title}: \texttt{u4lib}

\noindent
\textit{Licensing provisions}: MIT License.

\noindent
\textit{Code repository}: \url{https://github.com/Deep-SANE/u4lib}

\noindent
\textit{Programming language}: C++ (C++20), with Python bindings via pybind11.

\noindent
\textit{External libraries}: \texttt{u3lib} is a prerequisite, through which core utilities and other dependencies---namely 
\texttt{Eigen3} for QR-based null space solver, \texttt{Boost} for hash tables, 
\texttt{OpenMP} for multi-threading, \texttt{fmt} for formatting output, \texttt{libwignernj} for $\rm SU(2)$
coupling and recoupling coefficients, \texttt{KraNS} for sparse null-space solvers on large matrices (with a dense-QR fallback), 
and \texttt{pybind11} for Python binding---are inherited.

\noindent
\textit{Nature of problem}: Efficient, accurate, and reproducible computation of 
Wigner reduced coupling coefficients for the two subgroup chains
$\mathrm{U(4)}\supset\mathrm{U(3)}$ and
$\mathrm{SU(4)}\supset\mathrm{SU_S(2)}\otimes\mathrm{SU_T(2)}$, 
Racah recoupling coefficients (6-$\rm U(4)$ and 9-$\rm U(4)$ coefficients), 
and the associated representation-theoretic quantities
(dimensions, Casimir invariants, multiplicities, basis enumeration, generator matrix
elements, and basis transformations) necessitated by symmetry-adapted
nuclear-structure calculations and similar applications.

\noindent
\textit{Solution method}: The reduced coupling and recoupling coefficients are built from
the highest-weight canonical $\rm U(4) \supset U(3) \supset U(2) \supset U(1)$ full coupling coefficients, 
obtained as the orthonormal null space of the canonical raising generators, and
propagated by the lowering counterparts as shown in \cite{Dang2024EPJP2}; the physical
$\rm SU(4) \supset \mathrm{SU_S(2)}\otimes\mathrm{SU_T(2)}$ coefficients follow from the
physical-to-canonical basis transformation, which is also determined through the null space of the
spin raising $S_+$ and isospin raising $T_+$ operators \cite{Dang2024EPJP2}.

\section{Introduction}
\label{sec:intro}

Wigner's supermultiplet symmetry was proposed in 1937 as the overarching scheme that
unifies the spin and isospin degrees of freedom of atomic nuclei \cite{Wigner1937PR}.
It is not an exact symmetry of the strong interaction, but rather emerges in the limit of many colors
in QCD \cite{Kaplan1996PLB} and at a nuclear-force resolution of roughly 500~MeV \cite{Lee2021PRL,Timoteo2012PRC}.
In spite of its broad range of applications, from nuclear masses \cite{Cauvin1981NPA,Isacker1997FoP,Dang2026_2},
$\beta$-systematics \cite{Wigner1939PR,Fujita1965NP,Muli2026PRL}, 
and Gamow-Teller resonances \cite{Lutostansky2016EPJ},
to more recent lattice simulations \cite{Lee2025ARNPS} and algebraic models that
mitigate symmetry breaking in heavy nuclei \cite{Isacker1999PRL,Kota2024PScr}, there has been 
limited effort to probe and leverage the supermultiplet symmetry in first-principles shell-model
calculations as compared to its spatial $\rm SU(3)$ \cite{Elliott1958PRSA,Elliott1958PRSA-II,Dytrych2013PRL}
and symplectic $\rm Sp(3,R)$ \cite{Dytrych2020PRL,McCoy2020PRL} counterparts.

This underdevelopment stems from the fact that working in the physical spin-isospin 
$\rm SU(4) \supset SU_S(2) \otimes SU_T(2)$ basis 
requires $\rm U(4)$/$\rm SU(4)$ coupling and recoupling coefficients. 
While the computation of these coefficients was studied earlier by Hecht and
Pang \cite{Hecht1969JMP}, by Draayer \cite{Draayer1970JMP}, by Partensky and Maguin \cite{Partensky1978JMP}
and by Rowe and Repka \cite{Rowe1997FoP}, no method produced scalable software
for large-scale nuclear structure calculations.
A recent Python package \cite{Quintero2026CPC} that automates the evaluation of 
$\rm SU(4) \supset SU_S(2) \otimes SU_T(2)$ branching multiplicity has been advanced.
Most recently, building on efforts by Pan \textit{et al.} to develop the physical basis and 
elementary physical coupling coefficients \cite{Pan2023EPJP,Pan2023NPA,Pan2024CPC} 
along with previous work on canonical coupling coefficients for $\rm SU(4)$ \cite{Kuhn2008CPC} 
and the more general $\rm SU(N)$ \cite{Alex2011JMP}, Dang \textit{et al.} have developed
a generic algorithm \cite{Dang2024EPJP2} that supplies any and all physical 
$\rm SU(4) \supset SU_S(2) \otimes SU_T(2)$ coupling and recoupling coefficients,
propelling the first \textit{ab initio} shell-model calculations using Wigner's symmetry \cite{Dang2026_1}.

In a nutshell, the new algorithm leverages a property of the group generators:
the highest-weight state of an irrep is annihilated by the raising operators,
which impose a system of homogeneous linear equations that can be solved
by a numerical null space solver
to yield the full coupling coefficients for the highest-weight state as an orthonormal set of
null space vectors. For the large sparse matrices arising in this approach,
a dedicated iterative null-space solver, now known as
\texttt{KraNS}, was developed \cite{Krastenov2026MSc}.
All other coefficients then naturally follow through the action
of the lowering operators. The same strategy has also been employed to modernize
the 1973 Draayer-Akiyama formalism \cite{Draayer1973JMP,Akiyama1973CPC}
for computing $\rm SU(3)$ coupling coefficients \cite{Dang2024EPJP1},
implemented in the modern library \texttt{u3lib}~\cite{Langr2026u3lib}.

This paper presents \texttt{u4lib}---a header-only C++ library with Python wrappers 
and a sibling of \texttt{u3lib}---which implements the algorithm demonstrated in
Ref.~\cite{Dang2024EPJP2} and is part of a broader effort called \texttt{DeepSANE} 
(\textbf{Deep}-learning and \textbf{S}ymmetry-adapted \textbf{A}b-initio \textbf{N}uclear 
modeling: an AI-ready \textbf{E}cosystem\footnote{See the \texttt{DeepSANE} GitHub organization: 
\url{https://github.com/Deep-SANE}}). To our knowledge, this is the first library capable of
computing generic coupling and recoupling coefficients for the supermultiplet symmetry.
A calculation herein can begin with  either a canonical state in the 
$\rm U(4) \supset U(3) \supset U(2) \supset U(1)$ chain 
or a physical state labeled by spin and isospin quantum numbers in the
$\rm SU(4) \supset SU_S(2)\otimes SU_T(2)$ chain. The library supplies the
transformation between these two bases, state conjugation, generator matrix elements,
Kronecker (tensor-product) decompositions, and Casimir invariants alongside the
coupling and recoupling routines. Its dependence on \texttt{u3lib} reflects the canonical
subgroup chain and allows the two libraries to share numerical infrastructure.

Following the broader effort to modernize scientific workflows in the agentic-AI era,
we also provide agent-ready domain skills in \texttt{u4lib}. With an aim similar to that of
the Genesis Mission's agent skill catalog\footnote{See the DOE Genesis
Mission skill catalog: \url{https://gitlab.osti.gov/genesis/genesis-skills}.} of 
the U.S. Department of Energy, these skills empower AI assistants to become
specialists capable of writing programs, debugging, and facilitating adoption.
To demonstrate their usefulness, we assess agents' performance through three coding tasks, 
measuring whether access to the skill catalog changes correctness, token use, runtime, and API cost.

Note that while our work lies primarily within nuclear physics, the same algebraic structure is also
adopted in various fields. In quantum information, the two-qubit
unitary group is $\rm SU(4)$, and the geometry of its nonlocal gates follows from
a Cartan decomposition of the algebra \cite{Zhang2003PRA}. In ultracold-atom
physics, alkaline-earth atoms realize $\rm SU(N)$-symmetric magnetism that
includes $\rm SU(4)$ \cite{Gorshkov2010NP}, while in condensed matter the
spin-valley degrees of freedom of graphene form an approximate $\rm SU(4)$
quantum Hall ferromagnet \cite{Young2012NP}. These applications may use the 
representation data supplied by \texttt{u4lib}, though their physical models 
and basis conventions must be handled separately---we cordially invite researchers
from other fields to contribute to the library to support their respective applications.

The rest of the paper is organized as follows.
Sections~\ref{sec:structure} and \ref{sec:classes} describe the source layout
and C++ reference interface. Section~\ref{sec:python} covers Python use, and
Sec.~\ref{sec:agents} explains the skills and reports the coding experiments.
Build instructions appear in Sec.~\ref{sec:build}. Section~\ref{sec:summary}
summarizes the article.

\section{Library structure}
\label{sec:structure}

The headers contain the U(4) implementation and use \texttt{u3lib} as a
dependency. Examples, bindings, tests, and documentation occupy separate
directories:
\begin{itemize}
    \item \verb|include/|: the library itself (all algorithms live here; user
    programs include these headers directly):
    \begin{itemize}
        \item \verb|U4irrep.h|: the irrep classes \verb|su4::irrep|, \verb|u4::irrep|;
        \item \verb|U4gelfand.h|, \verb|internal/gelfand/|: the Gelfand state class \verb|u4::gelfand|;
        \item \verb|U4basis.h|: basis enumeration (\verb|gelfandBasis|, \verb|physicalBasis|), the inner multiplicity \verb|Zeta|, the \verb|u4::physicalvector| class, and the generator-action predicates;
        \item \verb|U4decomposition.h|: the outer multiplicity \verb|Eta| and the tensor-product \verb|decomposition|;
        \item \verb|U4U3xU4U3.h|, \verb|U4SU2SU2xU4SU2SU2.h|: the canonical and physical coupling coefficients and their databases;
        \item \verb|U4_U6Z6Coeffs.h|, \verb|U4_U9Coeffs.h|: the 6-$\rm U(4)$ and 9-$\rm U(4)$ recoupling coefficients and their databases;
        \item \verb|U4SU2SU2_to_U4U3U2U1.h|: the physical-to-canonical transformation;
        \item \verb|U4generatorsRME.h|: the physical-generator reduced matrix elements \verb|RME_S|, \verb|RME_T|, \verb|RME_U|;
        \item \verb|Definitions.h|: the structured coefficient containers;
        \item \verb|u4lib.h|: the umbrella header and the \verb|initialize|/\verb|finalize| lifecycle.
    \end{itemize}
    \item \verb|src/|: small example driver programs (one \verb|.cpp| per task) that call the headers and print results; compiled to \verb|bin/|.
    \item \verb|python/|: the pybind11 extension (\verb|pyu4lib.cpp| $\to$ \verb|u4lib.so|), its build (\verb|make python|, \verb|build.sh|), build notes (\verb|BUILD_NOTES.md|), and a tutorial Jupyter notebook (\verb|tutorial.ipynb|).
    \item \verb|agents/|: the AI-agent skill catalog: an index (\verb|AGENTS.md|), shared runtime instructions, and one skill per capability.
    \item \verb|test/|: the regression suite (\verb|test_*.cpp|, the \verb|*_cases.h| case sets, and \verb|reference/| documented values), run by \verb|make test|.
    \item \verb|doc/|: this manual (LaTeX source).
    \item \verb|bin/|: build output (not source).
\end{itemize}

\section{Classes and library interface}
\label{sec:classes}

The interface separates algebraic objects from coefficient databases.
Irreps, basis states, and their methods can be used immediately; database
queries require the lifecycle calls below. U(4) types use the \verb|u4|
namespace and inherit SU(4) functionality from \verb|su4|.
Table~\ref{tab:main-funcs} identifies the relevant header for each operation.

\begin{table}[h!]
    \centering
    \begin{tabular}{p{4.3cm}p{5.9cm}p{6.5cm}}
    \hline\hline
         Header & Entry point & Task \\
        \hline
         \verb|U4irrep.h| & \verb|dim|, \verb|C1/C2/C3/C4|, \verb|conjugate| & irrep dimension, Casimirs, conjugation \\
         \verb|U4basis.h| & \verb|Zeta|, \verb|gelfandBasis|, \verb|physicalBasis| & inner multiplicity; basis enumeration \\
         \verb|U4decomposition.h| & \verb|Eta|, \verb|decomposition| & outer multiplicity; Kronecker products \\
         \verb|U4U3xU4U3.h| & \verb|wigU4U3db| & $\mathrm{U(4)}{\supset}\mathrm{U(3)}$ canonical coupling coefficients \\
         \verb|U4SU2SU2xU4SU2SU2.h| & \verb|wigU4SU2SU2db| & $\mathrm{SU_S(2)}{\times}\mathrm{SU_T(2)}$ physical coupling coefficients \\
         \verb|U4_U6Z6Coeffs.h| & \verb|u4u6db|, \verb|u4z6db| & 6-$\rm U(4)$ recoupling coefficients \\
         \verb|U4_U9Coeffs.h| & \verb|u4u9db| & 9-$\rm U(4)$ recoupling coefficients \\
         \verb|U4SU2SU2_to_U4U3U2U1.h| & \verb|phys_to_cano| & physical-to-canonical transformation \\
         \verb|U4generatorsRME.h| & \verb|RME_S|, \verb|RME_T|, \verb|RME_U| & reduced matrix elements of $\mathbf{S}$, $\mathbf{T}$, $\mathbf{U}$ \\
    \hline\hline
    \end{tabular}
    \caption{Main entry points of \texttt{u4lib}.}
    \label{tab:main-funcs}
\end{table}

\subsection{Lifecycle}
In \verb|u4lib.h|:
\begin{itemize}
    \item \verb|initialize()| / \verb|initialize(&argc,&argv)|: set up the caches of SU(2) coupling and recoupling coefficients, and the U(3) and U(4) coefficient databases (the cascade descends into \texttt{u3lib});
    \item \verb|finalize()|: release these resources in reverse order;
    \item \verb|initialize_thread()| / \verb|finalize_thread()|: per-thread setup and teardown for multithreaded use;
    \item \verb|clear()|: discard cached coefficients while leaving the databases intact;
    \item \verb|is_initialized()|: report the current state.
\end{itemize}
Shared initialization: \texttt{u4lib} initializes its \texttt{u3lib}
dependency automatically. If the application also calls \texttt{u3lib}
directly, initialize both libraries at startup and defer finalization until
all coefficient calculations have finished. Their \texttt{libwignernj} backend is shared;
finalizing either library while the other still needs it causes runtime errors.

\subsection{Irreps: \texttt{su4::irrep} and \texttt{u4::irrep}}
(\verb|U4irrep.h|.) \verb|su4::irrep| stores an SU(4) irrep $(P_1,P_2,P_3)$;
\verb|u4::irrep| derives from it and adds the fourth label $n_{44}$ to fix the total quanta.

\textbf{Constructors:}
\begin{itemize}
    \item \verb|su4::irrep()|: default (empty) SU(4) irrep.
    \item \verb|su4::irrep(P1,P2,P3)|: SU(4) irrep from its labels.
    \item \verb|u4::irrep()|: default (empty) U(4) irrep.
    \item \verb|u4::irrep(n14,n24,n34,n44)|: U(4) irrep from its four labels.
    \item \verb|u4::irrep(P1,P2,P3)|: normalized representative.
    \item \verb|u4::irrep(su4::irrep, A)|: fix the total quanta $A$.
\end{itemize}

\textbf{Label getters:}
\begin{itemize}
    \item \verb|P1()|, \verb|P2()|, \verb|P3()|: the SU(4) labels.
    \item \verb|n14()|, \verb|n24()|, \verb|n34()|, \verb|n44()|: the four U(4) labels (only for \verb|u4::irrep|).
\end{itemize}

\textbf{Dimension and invariants:}
\begin{itemize}
    \item \verb|dim()|: the irrep dimension (both classes).
    \item \verb|C2()|, \verb|C3()|, \verb|C4()|: the SU(4) Casimirs of orders 2, 3, 4 (both classes, shift-invariant).
    \item \verb|C1()|: the first-order Casimir $=n_{14}{+}n_{24}{+}n_{34}{+}n_{44}$ (total quanta), on \verb|u4::irrep| only (SU(4) is U(4) with the U(1) trace generator removed, so \verb|su4::irrep| has no \verb|C1()|).
    \item \verb|quanta()|: synonym of \verb|C1()|.
\end{itemize}

\textbf{Other:}
\begin{itemize}
    \item \verb|zero()|: test for the trivial irrep.
    \item \verb|conjugate()|: the conjugate irrep $(P_3,P_2,P_1)$.
    \item \verb|conjugate_phase_power()|: the conjugation phase factor.
    \item \verb|normalize()|: set $n_{44}=0$ for \verb|u4::irrep|.
    \item comparison operators and stream output.
    \item \verb|writeBinary| / \verb|readBinary|: serialization for binary files.
\end{itemize}

\subsection{Gelfand states: \texttt{u4::gelfand}}
(\verb|U4gelfand.h|.) The canonical basis state with U(4) labels
$[n_{14},n_{24},n_{34},n_{44}]$ and U(3)$\supset$U(2)$\supset$U(1) sublabels
$[n_{13},n_{23},n_{33}]$, $[n_{12},n_{22}]$, $n_{11}$.

\textbf{Constructor:}
\begin{itemize}
    \item \verb|u4::gelfand(n14,n24,n34,n44,n13,n23,n33,n12,n22,n11)|: build a Gelfand state from its labels.
\end{itemize}

\textbf{Label getters and validity check:}
\begin{itemize}
    \item \verb|n14()| \dots\ \verb|n11()|: the ten Gelfand-pattern labels.
    \item \verb|check()|: test the betweenness (interlacing) condition.
    \item \verb|normalize()|: shift all quantum numbers by $n_{44}$.
\end{itemize}

\textbf{Subgroup labels:}
\begin{itemize}
    \item \verb|P1()|, \verb|P2()|, \verb|P3()|: the SU(4) labels.
    \item \verb|Lambdax2()|: $2\times$ the embedded SU(2) angular momentum.
    \item \verb|MLambdax2()|: $2\times$ its projection.
\end{itemize}

\textbf{Conjugation:}
\begin{itemize}
    \item \verb|conjugate()|: the Gelfand state's conjugate.
    \item \verb|conjugate_phase()|: the accompanying conjugation phase.
\end{itemize}

\textbf{Generator matrix elements:} Each method returns the matrix element of a
U(4) canonical generator $E_{ij}$ acting on a Gelfand state. The suffix identifies
the target branch: \verb|_12|/\verb|_22| selects $n_{12}$ or $n_{22}$, while
\verb|_13|/\verb|_23|/\verb|_33| selects $n_{13}$, $n_{23}$, or $n_{33}$.
For generators that change both rows, the two suffixes specify the corresponding
pair of labels (see Ref.~\cite{Dang2024EPJP2}).
\begin{itemize}
    \item \verb|e11()|, \verb|e22()|, \verb|e33()|, \verb|e44()|: the diagonal weight generators.
    \item \verb|e12()|: the $E_{12}$ element.
    \item \verb|e21()|: the $E_{21}$ element.
    \item \verb|e23_12()|, \verb|e23_22()|: the $E_{23}$ element, per target branch.
    \item \verb|e32_12()|, \verb|e32_22()|: the $E_{32}$ element, per target branch.
    \item \verb|e13_12()|, \verb|e13_22()|: the $E_{13}$ element, per target branch.
    \item \verb|e31_12()|, \verb|e31_22()|: the $E_{31}$ element, per target branch.
    \item \verb|e34_13()|, \verb|e34_23()|, \verb|e34_33()|: the $E_{34}$ element, per target branch.
    \item \verb|e43_13()|, \verb|e43_23()|, \verb|e43_33()|: the $E_{43}$ element, per target branch.
    \item \verb|e24_13_12()|, \verb|e24_13_22()|, \verb|e24_23_12()|, \verb|e24_23_22()|, \verb|e24_33_12()|, \verb|e24_33_22()|: the $E_{24}$ element, per target branch.
    \item \verb|e42_13_12()|, \verb|e42_13_22()|, \verb|e42_23_12()|, \verb|e42_23_22()|, \verb|e42_33_12()|, \verb|e42_33_22()|: the $E_{42}$ element, per target branch.
    \item \verb|e14_13_12()|, \verb|e14_13_22()|, \verb|e14_23_12()|, \verb|e14_23_22()|, \verb|e14_33_12()|, \verb|e14_33_22()|: the $E_{14}$ element, per target branch.
    \item \verb|e41_13_12()|, \verb|e41_13_22()|, \verb|e41_23_12()|, \verb|e41_23_22()|, \verb|e41_33_12()|, \verb|e41_33_22()|: the $E_{41}$ element, per target branch.
\end{itemize}
The raising-action predicates in \verb|U4basis.h| use the corresponding
uppercase names: \verb|E12|, \verb|E23_12|, \verb|E23_22|,
\verb|E13_12|, \verb|E13_22|, \verb|E34_13|, \verb|E34_23|,
\verb|E34_33|, and the six branches each of \verb|E24| and \verb|E14|
with the same paired suffixes listed above. They test whether a candidate final
state has the label changes required by that raising-generator branch.

\subsection{Physical states: \texttt{u4::physicalvector}}
(\verb|U4basis.h|.) The $\mathrm{SU_S(2)}\times\mathrm{SU_T(2)}$ reduced spin-isospin
state $\ket{(P_1 P_2 P_3)\,\zeta\,S\,T}$, where spin and isospin are stored
doubled as $SS=2S$, $TT=2T$, and $\zeta$ is the inner multiplicity.

\textbf{Constructor:}
\begin{itemize}
    \item \verb|u4::physicalvector(irrep, SS, TT)|: a physical state of doubled spin $SS=2S$ and isospin $TT=2T$.
\end{itemize}

\textbf{Accessors:}
\begin{itemize}
    \item \verb|irr()|: the underlying irrep.
    \item \verb|P1()|, \verb|P2()|, \verb|P3()|: the SU(4) labels.
    \item \verb|n14()| \dots\ \verb|n44()|: the U(4) labels.
    \item \verb|SS()|, \verb|TT()|: the doubled spin and isospin.
    \item \verb|zeta()| / \verb|zetamax()|: the inner multiplicity of $(S,T)$.
\end{itemize}

\subsection{Free functions}
\textbf{Multiplicities:}
\begin{itemize}
    \item \verb|u4::Zeta(irrep,SS,TT)|: inner multiplicity $\zeta_{\max}$ of $(S,T)$.
    \item \verb|u4::Eta(A,B,C)|: outer multiplicity $\eta_{\max}$ of irrep $C$ in the $A\otimes B$ product (\verb|U4decomposition.h|).
\end{itemize}

\textbf{Bases:}
\begin{itemize}
    \item \verb|u4::gelfandBasis(irrep)|: enumerate every Gelfand state of the irrep.
    \item \verb|u4::physicalBasis(irrep)|: list all physical states with nonzero $\zeta$.
    \item \verb|u4::findGelfand(g,gArr)|: find the position of a state inside the \verb|gArr| Gelfand basis.
\end{itemize}

\textbf{Decomposition:}
\begin{itemize}
    \item \verb|u4::decomposition(A,B)|: the complete Kronecker series, as a (list of $C$, list of $\eta_{\max}$) pair satisfying $\dim A\dim B=\sum_C\eta_{\max}(C)\dim C$.
\end{itemize}

\textbf{Physical generator reduced matrix elements:} Each takes two
\verb|physicalvector| states (bra, ket) and two \verb|std::vector<int>|
out-parameters, and returns an \verb|Eigen::MatrixXd| of shape
$\zeta'_{\max}\times\zeta_{\max}$ (\verb|U4generatorsRME.h|).
\begin{itemize}
    \item \verb|RME_S|, \verb|RME_T|: diagonal in $(S,T)$, with $\langle\|\mathbf{S}\|\rangle=\sqrt{S(S{+}1)(2S{+}1)(2T{+}1)}$ and $\langle\|\mathbf{T}\|\rangle=\sqrt{T(T{+}1)(2T{+}1)(2S{+}1)}$.
    \item \verb|RME_U|: the rank-$(1,1)$ Gamow--Teller-type tensor ($\Delta S,\Delta T\in\{-1,0,+1\}$).
\end{itemize}

\textbf{Transformation:} Each routine expands the physical state \verb|p| (at
doubled projections \texttt{MMS} and \texttt{MMT}) over the Gelfand basis \verb|gArr|, returning the
canonical-state indices together with an amplitude matrix whose columns are
orthonormal, one per $\zeta$.
\begin{itemize}
    \item \verb|u4::phys_to_cano(p,MMS,MMT,gArr)|: the expansion at a single projection $(M_S,M_T)$.
    \item \verb|u4::phys_to_cano_allM(p,gArr)|: the expansion over all projections.
\end{itemize}

\subsection{Coefficient containers}
(\verb|Definitions.h|.) Coefficients are grouped into containers whose axes
correspond to multiplicity labels. Each container provides coordinate access
through \verb|operator()|, flat indexing through \verb|operator[]|, and
formatted output through \verb|print()| or \verb|operator<<|:
\begin{itemize}
    \item \verb|block(i0,...,id)| selects a coefficient by its multiplicity indices,
    in the order given below. Indices are zero-based and must be smaller than
    their corresponding \verb|*max()| counts.
    \item \verb|block[i]| accesses the same data with a single index,
    $0\leq i<\verb|block.size()|$. Storage is row-major: the last axis varies
    fastest. For axis sizes $d_0,\ldots,d_{m-1}$, the flat index is
    $i=\sum_{j=0}^{m-1}i_j\prod_{k=j+1}^{m-1}d_k$, with an empty product equal
    to one. For four axes, $i=((i_0d_1+i_1)d_2+i_2)d_3+i_3$.
    \item \verb|data()| returns a const reference to the underlying storage
    object; \verb|size()| gives the number of coefficients, and \verb|empty()|
    indicates whether the block contains any coefficient.
    \item \verb|print(s)| and \verb|s << block| print the multiplicity indices
    and coefficient values to stream \verb|s|. With no argument,
    \verb|print()| writes to \verb|std::cout|.
\end{itemize}
For containers with writable storage, non-const indexing returns a writable
reference, while const indexing returns the coefficient by value. The cached
\verb|U4U9| database result uses a read-only view; access it through a const
container. Storage ownership is discussed at the end of this subsection.

In the notation below, $A,B,C,D$ are U(4) irreps. Labels such as $AB,BC,AC,BD,CD$
refer to selected intermediate irreps, not to the tensor products themselves.
We use $R$ for the final irrep in recoupling.
The outer multiplicity $\eta$ distinguishes occurrences of a U(4) irrep in a
product; $\rho$ plays the corresponding role for the U(3) subgroup coupling.
The inner multiplicity $\zeta$ distinguishes repeated spin-isospin pairs within
an SU(4) irrep. All multiplicity indices below follow the library's zero-based
convention.

\textbf{\texttt{U4U3}}: canonical coupling of $A\otimes B\to C$, reduced with
respect to U(3):
$
  \left\langle A a; B b \middle\| C c\right\rangle_{\rho,\eta},
$
where $a,b,c$ are fixed U(3) sublabels $[n_{13},n_{23},n_{33}]$.
Unlike the $\rm U(3) \supset U(2)$ case, the subgroup coupling itself can have outer
multiplicity. The block therefore has two axes, $(\rho,\eta)$.
\begin{itemize}
    \item \verb|rhomax()|: multiplicity of the U(3) coupling $a\otimes b\to c$.
    \item \verb|etamax()|: multiplicity of the U(4) coupling $A\otimes B\to C$.
    \item \verb|operator()(rho,eta)|: selects a coefficient with flat index
    $\rho\,\eta_{\max}+\eta$.
    \item \verb|print(s = std::cout)|: prints the coefficients over all $(\rho,\eta)$ pairs.
\end{itemize}

\textbf{\texttt{U4SU2SU2}}: physical coupling of $A\otimes B\to C$, reduced
with respect to $\mathrm{SU_S(2)}\otimes\mathrm{SU_T(2)}$:
$
  \left\langle A\,\zeta_A S_A T_A;\,B\,\zeta_B S_B T_B
  \middle\| C\,\zeta_C S_C T_C\right\rangle_{\eta}.
$
The spins and isospins are fixed for each block and supplied to the library in
doubled form, $SS=2S$ and $TT=2T$. The three inner multiplicities and the U(4)
outer multiplicity form the four axes.
\begin{itemize}
    \item \verb|zetaAmax()|: multiplicity of $(S_A,T_A)$ in $A$.
    \item \verb|zetaBmax()|: multiplicity of $(S_B,T_B)$ in $B$.
    \item \verb|zetaCmax()|: multiplicity of $(S_C,T_C)$ in $C$.
    \item \verb|etamax()|: multiplicity of $C$ in $A\otimes B$.
    \item \verb|operator()(zetaA,zetaB,zetaC,eta)|.
    \item \verb|print(s = std::cout)|: prints the coefficients over
    $(\zeta_A,\zeta_B,\zeta_C,\eta)$.
\end{itemize}

\textbf{\texttt{U4U6}}: $U$-recoupling between $(AB)C$ and $A(BC)$:
$
 U\!\left(A,B,R,C;\,AB,\eta_{AB},\eta_{AB\_C};\,
 BC,\eta_{BC},\eta_{A\_BC}\right).
$
The intermediate irreps $AB$ and $BC$ are fixed; the four axes resolve the
multiplicities along the two coupling routes.
\begin{itemize}
    \item \verb|etaABmax()|: multiplicity of $A\otimes B\to AB$.
    \item \verb|etaAB_Cmax()|: multiplicity of $AB\otimes C\to R$.
    \item \verb|etaBCmax()|: multiplicity of $B\otimes C\to BC$.
    \item \verb|etaA_BCmax()|: multiplicity of $A\otimes BC\to R$.
    \item \verb|operator()(etaAB,etaAB_C,etaBC,etaA_BC)|.
    \item \verb|print(s = std::cout)|: prints the coefficients over
    $(\eta_{AB},\eta_{AB\_C},\eta_{BC},\eta_{A\_BC})$.
\end{itemize}

\textbf{\texttt{U4Z6}}: $Z$-recoupling between $(AB)C$ and $(AC)B$:
$
 Z\!\left(B,A,R,C;\,AB,\eta_{AB},\eta_{AB\_C};\,
 AC,\eta_{AC},\eta_{AC\_B}\right).
$
The intermediate irreps $AB$ and $AC$ are fixed; the four axes resolve the
multiplicities along the two coupling routes.
\begin{itemize}
    \item \verb|etaABmax()|: multiplicity of $A\otimes B\to AB$.
    \item \verb|etaAB_Cmax()|: multiplicity of $AB\otimes C\to R$.
    \item \verb|etaACmax()|: multiplicity of $A\otimes C\to AC$.
    \item \verb|etaAC_Bmax()|: multiplicity of $AC\otimes B\to R$.
    \item \verb|operator()(etaAB,etaAB_C,etaAC,etaAC_B)|.
    \item \verb|print(s = std::cout)|: prints the coefficients over
    $(\eta_{AB},\eta_{AB\_C},\eta_{AC},\eta_{AC\_B})$.
\end{itemize}

\textbf{\texttt{U4U9}}: 9-U(4) recoupling between $(AB)(CD)$ and $(AC)(BD)$:
\begin{equation}
 \left[\begin{array}{cccc}
 A & B & AB & \eta_{AB} \\
 C & D & CD & \eta_{CD} \\
 AC & BD & R & \eta_{AC\_BD} \\
 \eta_{AC} & \eta_{BD} & \eta_{AB\_CD} &
 \end{array}\right].
\end{equation}
The last column labels the row couplings, and the last row labels the column
couplings. In the code, \verb|etaABCD| denotes $\eta_{AB\_CD}$ for
$AB\otimes CD\to R$, whereas \verb|etaACBD| denotes $\eta_{AC\_BD}$ for
$AC\otimes BD\to R$. The storage order is
$(\eta_{AC},\eta_{BD},\eta_{AB\_CD},\eta_{AB},\eta_{CD},\eta_{AC\_BD})$.
\begin{itemize}
    \item \verb|etaACmax()|: multiplicity of $A\otimes C\to AC$.
    \item \verb|etaBDmax()|: multiplicity of $B\otimes D\to BD$.
    \item \verb|etaABCDmax()|: multiplicity of $AB\otimes CD\to R$.
    \item \verb|etaABmax()|: multiplicity of $A\otimes B\to AB$.
    \item \verb|etaCDmax()|: multiplicity of $C\otimes D\to CD$.
    \item \verb|etaACBDmax()|: multiplicity of $AC\otimes BD\to R$.
    \item \verb|operator()(etaAC,etaBD,etaABCD,etaAB,etaCD,etaACBD)|.
    \item \verb|print(s = std::cout)|: prints all six multiplicity indices
    and the corresponding coefficients.
\end{itemize}

The classes are templated on their storage type. The first four database
containers own their coefficient vectors. With \verb|U4LIB_ENABLE_U4U9_CACHE| on,
\verb|get_U4U9| instead returns a container holding a
\verb|std::span<const double>| into a thread-private scratch buffer. A later
query can overwrite that buffer, so copy the coefficients into owning storage
before retaining results across calls; copying the view alone does not retain
the data. Otherwise, the returned container owns a
\verb|std::vector<double>|.

\subsection{Coefficient databases}
Every coupling and recoupling coefficient is evaluated on first use and memoized
inside a process-wide singleton database: \verb|u4::wigU4U3db()|,
\verb|u4::wigU4SU2SU2db()|, \verb|u4::u4u6db()|, \verb|u4::u4z6db()|, and
\verb|u4::u4u9db()|. Each database offers two complementary retrieval styles:
\begin{itemize}
    \item \textbf{Structured container getters} (\verb|get_U4U3|, \verb|get_U4SU2SU2|, \verb|get_U4U6|, \verb|get_U4Z6|, \verb|get_U4U9|): hand back the typed containers described above (\verb|U4U3|, \verb|U4SU2SU2|, \verb|U4U6|, \verb|U4Z6|, \verb|U4U9|), with coordinate access through \verb|operator()|, flat access through \verb|operator[]|, the \verb|*max()| shape queries, and stream printing.
    \item \textbf{Direct buffer getters} (\verb|get|): return the coefficients as a bare buffer, bypassing the container wrapper. The three recoupling databases add a raw-integer overload of \verb|get| that consumes the partition labels directly and derives the intermediate multiplicities on its own.
\end{itemize}

Taking the databases in turn:
\begin{itemize}
    \item \textbf{Canonical coupling database} (\verb|u4::wigU4U3db()|):
    \begin{itemize}
        \item \verb|get_U4U3(| $n_{14}^a,n_{24}^a,n_{34}^a,n_{44}^a,n_{13}^a,n_{23}^a,n_{33}^a,\ \dots^{b},\ \dots^{c}$ \verb|)| $\to$ \verb|U4U3|: the $\mathrm{U(4)}\supset\mathrm{U(3)}$ reduced coupling indexed by $(\rho,\eta)$. 
        \item \verb|get(| $n_{14}^a,n_{24}^a,n_{34}^a,n_{44}^a,n_{13}^a,n_{23}^a,n_{33}^a,\ \dots^{b},\ \dots^{c}$ \verb|)| $\to$ \verb|Eigen::MatrixXd|: the same block as a $\rho_{\max}\times\eta_{\max}$ matrix; an empty matrix marks a forbidden coupling.
    \end{itemize}

    \item \textbf{Physical coupling database} (\verb|u4::wigU4SU2SU2db()|):
    \begin{itemize}
        \item \verb|get_U4SU2SU2(| $P_1^a,P_2^a,P_3^a,SS^a,TT^a,\ \dots^{b},\ \dots^{c}$ \verb|)| $\to$ \verb|U4SU2SU2|: the physical spin-isospin reduced coupling ($SS=2S$, $TT=2T$), indexed by $(\zeta_A,\zeta_B,\zeta_C,\eta)$.
        \item \verb|get(| $P_1^a,P_2^a,P_3^a,SS^a,TT^a,\ \dots^{b},\ \dots^{c}$ \verb|)| $\to$ \verb|std::vector<double>|: the same block flattened over the $(\zeta_A,\zeta_B,\zeta_C,\eta)$ axes.
    \end{itemize}

    \item \textbf{$U$-recoupling database} (\verb|u4::u4u6db()|):
    \begin{itemize}
        \item \verb|get_U4U6(| $A,B,ABC,C,AB,\ \eta_{AB},\eta_{AB\_C},\ BC,\eta_{BC},\eta_{A\_BC}$ \verb|)| $\to$ \verb|U4U6|: recouples $(AB)C\leftrightarrow A(BC)$, indexed by $(\eta_{AB},\eta_{AB\_C},\eta_{BC},\eta_{A\_BC})$.
        \item \verb|get(| $A,B,ABC,C,AB,\ \eta_{AB},\eta_{AB\_C},\ BC,\eta_{BC},\eta_{A\_BC}$ \verb|)| $\to$ \verb|std::vector<double>|: the flat form of length $\eta_{AB}\eta_{AB\_C}\eta_{BC}\eta_{A\_BC}$.
        \item \verb|get(| 24 integers \verb|)| $\to$ \verb|std::vector<double>|: a raw-integer overload taking the partition labels of $A,B,ABC,C,AB,BC$ and resolving the $\eta$ maxima internally.
    \end{itemize}

    \item \textbf{$Z$-recoupling database} (\verb|u4::u4z6db()|):
    \begin{itemize}
        \item \verb|get_U4Z6(| $B,A,ABC,C,AB,\ \eta_{AB},\eta_{AB\_C},\ AC,\eta_{AC},\eta_{AC\_B}$ \verb|)| $\to$ \verb|U4Z6|: recouples $(AB)C\leftrightarrow(AC)B$, indexed by $(\eta_{AB},\eta_{AB\_C},\eta_{AC},\eta_{AC\_B})$ (note the leading $B,A$ exchanges the first pair).
        \item \verb|get(| $B,A,ABC,C,AB,\ \eta_{AB},\eta_{AB\_C},\ AC,\eta_{AC},\eta_{AC\_B}$ \verb|)| $\to$ \verb|std::vector<double>|: the flat form of length $\eta_{AB}\eta_{AB\_C}\eta_{AC}\eta_{AC\_B}$.
        \item \verb|get(| 24 integers \verb|)| $\to$ \verb|std::vector<double>|: a raw-integer overload for the partition labels of $B,A,ABC,C,AB,AC$ and resolving the multiplicities internally.
    \end{itemize}

    \item \textbf{9-$\rm U(4)$ recoupling database} (\verb|u4::u4u9db()|):
    \begin{itemize}
        \item \verb|get_U4U9(| $A,B,C,D,AB,CD,AC,BD,ABCD,$ $\eta_{AC},\eta_{BD},\eta_{ABCD},\eta_{AB},\eta_{CD},\eta_{ACBD}$ \verb|)| $\to$ \verb|U4U9|: the coefficient container; its multiplicity arguments follow the container's axis order, the column channels $(\eta_{AC},\eta_{BD},\eta_{ABCD})$ ahead of the row channels $(\eta_{AB},\eta_{CD},\eta_{ACBD})$.
        \item \verb|get(| $A,B,C,D,AB,CD,AC,BD,ABCD,$ $\eta_{AB},\eta_{CD},\eta_{ABCD},\eta_{AC},\eta_{BD},\eta_{ACBD}$ \verb|)| $\to$ \verb|std::span<const double>| (or \verb|std::vector<double>|): the flat buffer, where the row channels $(\eta_{AB},\eta_{CD},\eta_{ABCD})$ instead come first, the reverse of \verb|get_U4U9|, matching the internal stride order.
        \item \verb|get(| 36 integers \verb|)| $\to$ \verb|std::vector<double>|: a raw-integer overload taking the partition labels of all nine participating irreps, where multiplicities are evaluated internally.
    \end{itemize}
\end{itemize}

The multiplicity counts passed to the recoupling getters must equal the
\verb|Eta| values of the matching channels, or an internal assertion fires. In
\verb|get_U4U9| the third count $\eta_{ABCD}$ belongs to the
$(AB)(CD)\to ABCD$ route and the sixth $\eta_{ACBD}$ to the
$(AC)(BD)\to ABCD$ route; interchanging them raises the same assertion. A
forbidden or vanishing coupling returns an empty container or buffer, which may
be queried for its shape but not indexed. Each database exposes a \verb|clear()|
method---and \verb|u4::clear()| empties all five at once---that drops the cached
blocks while keeping the allocated thread workspaces.

\subsection{Conventions and normalization}
\label{sub:conv}
Phase and normalization conventions follow Ref.~\cite{Dang2024EPJP2}; documented
values and the generation procedure are in \verb|test/reference/|.
\begin{itemize}
    \item \textbf{Doubled spin-isospin:} All spin and isospin labels are doubled: $SS=2S$, $TT=2T$ (and $MMS=2M_S$, $MMT=2M_T$). 
    \item \textbf{Multiplicity indexing:} The \verb|Zeta| and \verb|Eta| routines report the counts, and every container accessor is indexed from zero up to the corresponding maximum ($0,\dots,\text{max}{-}1$).
    \item \textbf{Canonical-coupling phase:} Within each $\eta$ block the highest-weight reduced coefficients set both phase and normalization. 
    When $\eta_{\max}>1$ the solver returns one admissible orthonormal null-space basis, and all coefficients carrying $\eta$ must use the same basis choice. Under conjugation a Gelfand state obeys $C\ket{g} = (-)^{n_{14}+n_{34}-n_{13}-n_{23}-n_{33}-n_{12}-n_{22}-n_{11}}\ket{g.\mathrm{conjugate}()}$ (\verb|gelfand::conjugate_phase()|).
    \item \textbf{Physical-to-canonical transformation:} A physical state $\ket{(P_1 P_2 P_3)\,\zeta\,S\,M_S\,T\,M_T}$ is expanded in canonical Gelfand states. The highest-weight state $\ket{(P_1 P_2 P_3)\,\zeta\,S\,S\,T\,T}$ is the null space of $S_+$ and $T_+$ \cite{Dang2024EPJP2}; $S_-$ and $T_-$ fix the lower-$M$ phases within a $\zeta$ multiplet. For $P_1\neq P_3$ the conjugation preserves $\zeta$,
    \begin{equation*}
      C\ket{(P_1 P_2 P_3)\,\zeta\,S\,M_S\,T\,M_T}=\epsilon\,(-)^{S-M_S+T-M_T}\,\ket{(P_3 P_2 P_1)\,\zeta\,S\,{-}M_S\,T\,{-}M_T},
      \quad
      \epsilon=\begin{cases}(-)^{P_1+P_2+P_3}, & P_1>P_3,\\[2pt] (-)^{P_2}, & P_1<P_3.\end{cases}
    \end{equation*}
For a self-conjugate irrep ($P_1=P_3$), conjugation instead acts as a projection-independent symmetric matrix $D$ ($D^2=\mathbb{1}$) that may mix $\zeta$. The direction-dependent $\epsilon$ reflects $C_{(P_3 P_2 P_1)}C_{(P_1 P_2 P_3)}=(-)^{P_1+P_3}\mathbb{1}$, which is $+1$ whenever $P_1+P_3$ is even and $-1$ otherwise; one may find an analogy in the $\mathrm{SU(2)}$ conjugation factor $C_j C_j = (-)^{2j}$.
    \item \textbf{Generator reduced matrix element factor:} \verb|RME_S|/\verb|RME_T|/\verb|RME_U| carry an extra $\sqrt{(2S'{+}1)(2T'{+}1)}$ factor.
    \item \textbf{Assertion checks.} Invalid labels or out-of-range indices abort via assertion.
\end{itemize}

\section{Python bindings}
\label{sec:python}

The \verb|u4lib| extension module reproduces the C++ API one-to-one. Build it
with \verb|make python| using the interpreter whose ABI matches the extension,
and keep \texttt{u3lib}'s \verb|python/| on the import path, since \verb|u4lib|
depends on it. The irreps (\verb|su4.irrep|, \verb|irrep|), \verb|gelfand|,
\verb|physicalvector|, the free functions \verb|Eta|, \verb|Zeta|,
\verb|decomposition|, \verb|gelfandBasis|, \verb|physicalBasis|, the generator
reduced matrix elements \verb|RME_S|/\verb|RME_T|/\verb|RME_U|, the transformations
\verb|phys_to_cano|/\verb|phys_to_cano_allM|, and the five database accessors all
carry the names and semantics of Sec.~\ref{sec:classes}. Each structured
container supports \verb|__call__| for multiplicity-indexed access, exposes its
flat \verb|data()| as a NumPy array, and prints itself; querying a database
before \verb|initialize()| or after \verb|finalize()| raises \verb|RuntimeError|. 
The \verb|U4U9| container, only a scratch-buffer view in
C++, is copied automatically so that the Python object owns its data.
The examples in the Jupyter notebook \verb|python/tutorial.ipynb| cover irreps, bases, coupling, and recoupling.
Consult \verb|python/BUILD_NOTES.md| for ABI and toolchain requirements.

\section{AI-ready interface}
\label{sec:agents}

\subsection{Skill catalog and orchestration}

When an user deploys an AI assistant with code-execution capability to use the library,
the top-level \texttt{README.md} instructs it to ignore the rest of the file and immediately
open the \verb|agents/| directory, which holds a catalog of skills that steer the
assistant through \texttt{u4lib} calculations. An index (\verb|agents/AGENTS.md|)
points to shared runtime instructions and one skill per capability, each documenting
the inputs, the exact library call, the relevant conventions, and a numerical
example. The agent follows these to drive the locally built library, and no network service is involved.

Setup is handled by the \verb|install| skill
(\verb|agents/install/SKILL.md|). It first checks whether the library is already
built, and otherwise runs \verb|make config|---which resolves every dependency
by inheriting \texttt{u3lib}'s toolchain and needs no manual environment
variables---followed by \verb|make| and \verb|make test| (a clean run prints
\verb|PASS: all test executables|); \verb|make python| adds the Python extension
when required. Once the build passes, the calculation skills become usable.

For a calculation, an AI assistant maps the request onto a skill, prepares the
inputs in the library's conventions, invokes the C++ or Python routine,
and returns the results. 
The numerical work stays inside
the library---the assistant neither derives nor approximates coefficients,
Casimirs, or matrix elements---and when a request is ambiguous or asks for an
undefined quantity it seeks clarification or explains the problem, e.g., the
first-order Casimir of an $\mathrm{SU(4)}$ irrep, which is not defined.

Table~\ref{tab:prompts} pairs each skill with a representative prompt whose
parenthetical answer was verified against the library. 
Spin and isospin inputs use the doubled labels $SS=2S$ and $TT=2T$.

\begin{table*}[tp]
    \centering
    \small
    \begin{tabular}{p{3.1cm}p{13.2cm}}
    \hline\hline
    Skill & Example prompt (expected result) \\
    \hline
    \verb|dim| & ``Dimension of the SU(4) irrep $(1,0,1)$ -- equivalently the U(4) irrep $[2,1,1,0]$?'' ($15$, the adjoint; shift-invariant) \\
    \verb|casimirs| & ``$C_2,C_3,C_4$ for the SU(4) adjoint $(1,0,1)$, and $C_1,C_2,C_3,C_4$ for the U(4) irrep $[2,1,1,0]$.'' (SU(4): $16,0,72$; U(4): $C_1{=}4,\,16,0,72$ -- $C_1$ is U(4)-only) \\
    \verb|cano-list| & ``Enumerate the canonical U(4) Gelfand basis of $(1,0,0)$ with each state's weights and subgroup labels.'' ($4$ states $={}$\,dim) \\
    \verb|phys-list| & ``List the physical spin-isospin basis $(\zeta,S,T)$ of $(2,0,0)$ and verify the dimension sum rule.'' ($(S,T){=}(1,1),(0,0)$, each $\zeta{=}1$; $\sum\zeta(2S{+}1)(2T{+}1){=}10$) \\
    \verb|cano-generators| & ``Act with $E_{12}$ on the Gelfand state $\ket{2,0,0,0;2,0,0;2,0;1}$ of $(2,0,0)$; give the resulting state and matrix element.'' ($\ket{2,0,0,0;2,0,0;2,0;2}$ with $\sqrt2$) \\
    \verb|phys-generators| & ``Compute $\langle(2,0,0)\,\zeta{=}1\,S{=}1\,T{=}1\|\mathbf{S}\|(2,0,0)\,\zeta{=}1\,S{=}1\,T{=}1\rangle$ and $\langle\|\mathbf{U}\|\rangle$ on the same state.'' ($\sqrt{18}\approx4.243$ and $3$) \\
    \verb|phys-to-cano| & ``Express $\ket{(2,0,0)\,\zeta{=}1\,S{=}1\,M_S{=}1\,T{=}1\,M_T{=}1}$ in the canonical Gelfand basis.'' (single state $\ket{2,0,0,0;2,0,0;2,0;2}$, amplitude $1$) \\
    \verb|inner-mult| & ``Spin-isospin multiplicity $\zeta$ of $(S,T){=}(1,1)$ in the adjoint $(1,0,1)$?'' ($1$) \\
    \verb|outer-mult| & ``Outer multiplicity $\eta$ of $(1,0,1)$ in $(1,0,1)\otimes(1,0,1)$?'' ($2$) \\
    \verb|decomposition| & ``Full decomposition of $(1,0,0)\otimes(1,0,0)$ with outer multiplicities; check the dimension sum rule.'' ($(2,0,0){+}(0,1,0)$, $\sum\eta\cdot\dim{=}16$) \\
    \verb|cano-coupling| & ``$\mathrm{SU(4)}{\supset}\mathrm{U(3)}$ reduced coupling for highest-weight states $[1,0,0,0]\,[1,0,0]\times[1,0,0,0]\,[1,0,0]\to[2,0,0,0]\,[2,0,0]$ of $(1,0,0)\times(1,0,0)\to(2,0,0)$; verify it equals $1$.'' \\
    \verb|phys-coupling| & ``Spin-isospin coupling $(1,0,0)\,\zeta{=}1\,S{=}\tfrac12\,T{=}\tfrac12 \times (1,0,0)\,\zeta{=}1\,S{=}\tfrac12\,T{=}\tfrac12 \to (2,0,0)\,\zeta{=}1\,S{=}1\,T{=}1$; verify it equals $1$.'' \\
    \verb|u6recoupling| & ``$U$-recoupling for $(1,0,0)\otimes(1,0,0)\otimes(1,0,0)\to(3,0,0)$ through intermediate irreps $AB=(2,0,0)$ and $BC=(2,0,0)$.'' ($1$) \\
    \verb|z6recoupling| & ``$Z$-recoupling between $(AB)C$ and $(AC)B$ routes for $(1,0,0)\otimes(1,0,0)\otimes(1,0,0)\to(3,0,0)$ with intermediate irreps $AB=(2,0,0)$ and $AC=(2,0,0)$.'' ($1$) \\
    \verb|u9recoupling| & ``$9$-coefficient recoupling for four fundamentals $(1,0,0)$ coupling to $(4,0,0)$ with intermediate irreps $AB=CD=AC=BD=(2,0,0)$.'' ($1$) \\
    \verb|conjugation| & ``Conjugate the SU(4) irrep $(2,0,0)$ and a Gelfand state of it, reporting the phases.'' (irrep $\to(0,0,2)$, phase power $3$) \\
    \hline\hline
    \end{tabular}
    \caption{The \texttt{u4lib} agent skills with a representative prompt for each.}
    \label{tab:prompts}
\end{table*}

\subsection{Benchmark objectives}
\label{app:benchmarking}
The benchmark tests whether the catalog helps an agent turn a U(4) calculation request
into a working program. Correctness is assessed separately from the resources
spent reaching a solution. We evaluate two hypotheses:
\begin{enumerate}
    \item {Resource efficiency:} Ready access to argument conventions and
    usage examples should reduce source-code searches and unsuccessful
    implementation attempts, lowering token use, runtime, and API cost.
    \item {Dependence on task complexity:} Guidance about memory ownership
    and thread safety should become more useful when the task moves beyond
    Python bindings to C++ storage management and parallel execution.
\end{enumerate}

\subsection{Experimental strategy}

\subsubsection{Benchmark Tasks}
The three tasks exercise different parts of the library's interface and place progressively greater demands on storage and concurrency:
\begin{itemize}
    \item {Task 1: Python Racah Recoupling Engine (\texttt{consumer/recouple.py}).}
    Write a Python utility using the pybind11 bindings to evaluate 6-$\rm U(4)$ recoupling coefficients via \verb|u4lib.u6recoupling| and \verb|u4lib.z6recoupling|. The implementation must check the unitary symmetry relations among fundamental representations (for instance, $(1,0,0) \otimes (1,0,0) \otimes (1,0,0) \to (3,0,0)$) and export structured numerical outputs. \emph{Architectural complexity: Low} (declarative Python bindings, automatic garbage collection, single-threaded execution).

    \item {Task 2: C++ 9-$\rm U(4)$ Batch Exporter (\texttt{consumer/u9\_batch.cpp}).}
    Build a C++ batch exporter for 9-$\mathrm{U}(4)$ recoupling blocks via \verb|u4::u4u9db()|. The program must parse integer input specifications for nine participating irreps, compute intermediate outer multiplicities $\eta$, retain coefficient buffers across requests without aliasing, configure build dependencies, and serialize block and entry records. \emph{Architectural complexity: Medium} (C++ typing, header inclusion, manual workspace lifecycle management, and partition label handling).

    \item {Task 3: C++ OpenMP Multi-threaded Concurrency (\texttt{consumer/coupling\_batch.cpp}).}
    Implement parallel canonical and physical coupling evaluations with OpenMP directives (\verb|#pragma omp parallel for|). The implementation must give each thread its own database cache and scratch state through per-thread initialization, guard shared output, set the required compiler flags (\verb|-fopenmp|), and scale across cores. \emph{Architectural complexity: High} (shared-memory concurrency, thread-private state isolation, race-hazard avoidance).
\end{itemize}

\subsubsection{Isolated workspace preparation}
Runs started in independent temporary directories, each with its own consumer build environment.
Files and build products from earlier trials were not reused. The workspace
snapshot determined what the agent could consult:
\begin{itemize}
    \item {Control (standard prompts, zero skills):} Source files, C++
    headers, and build scripts were available, but the \verb|agents/| directory was removed.
    \item {Treatment (domain skills):} The workspace also contained the
    repository's skill catalog (\verb|agents/skills/|), indexed by
    \verb|agents/AGENTS.md|. The prompt instructed the agent to consult it
    for conventions and usage patterns.
\end{itemize}

\subsubsection{Completion and validation}
Agents could read and edit files, inspect directories, and run shell commands
without intervention. A run ended when the agent declared completion or used
50 steps. The workspace was then frozen for evaluation against the hidden
tests. 
The grader compared numerical output with exact coupling and
recoupling values for several irreps, allowing a tolerance of $10^{-12}$.
We checked the grader itself 
including phase inversions, memory leaks, uninitialized reads, and race conditions. All such
validation cases were classified correctly: sensitivity and specificity were
100\%, with no false positives or negatives.

\subsubsection{Model and Statistical Protocol}
Every trial used Anthropic's Claude Sonnet 5.5 (\texttt{claude-sonnet-5-5}) in the same autonomous test environment. Each of the three tasks contributed $N = 10$ matched Control/Treatment pairs, giving 60 runs overall. For every run, four primary metrics were recorded:
\begin{enumerate}
    \item {Functional pass rate (\%):} Percentage of runs passing all hidden unit tests.
    \item {Total token consumption:} Cumulative sum of prompt tokens, cache-read tokens, and completion tokens consumed across all agent steps.
    \item {Wall-clock latency (s):} Total elapsed real time from task dispatch to final agent termination, measured externally by the evaluation harness.
    \item {Monetary cost (USD):} Total API billing incurred, computed from official token pricing.
\end{enumerate}
Differences were assessed using two-tailed Welch's $t$-tests (unequal variance assumption), reporting $t$-statistics, degrees of freedom, and $p$-values.

\begin{table}[tp]
    \centering
    \footnotesize
    \setlength{\tabcolsep}{2.8pt}
    \renewcommand{\arraystretch}{1.3}
    \caption{Performance of Control (standard prompts, zero skills) versus Treatment (repository-local domain skills) on \texttt{u4lib} with Claude Sonnet 5.5, over $N=10$ matched pairs per task (60 runs in total). Token counts, runtimes, and costs are given as $\text{Mean} \pm \text{Standard Deviation}$, and significance follows from two-tailed Welch's $t$-tests.}
    \label{tab:benchmark_results}
    \vspace{0.2cm}
    \begin{tabular}{llcccccc}
    \hline\hline
    Task & Condition & Pass Rate & Total Tokens & Time (s) & Cost (\$) & Token $\Delta$ & Time $\Delta$ \\
    \hline
    \textbf{Task 1: Python} 
      & Control ($N{=}10$)   & 90.0\% (9/10)  & $650,386 \pm 304,782$ & $97.35 \pm 42.34$  & \$0.3949 & --- & --- \\
      & Treatment ($N{=}10$) & 100\% (10/10)  & $318,884 \pm 109,257$ & $62.04 \pm 42.67$  & \$0.2377 & \textbf{$-$51.0\%}$^{**}$ & \textbf{$-$36.3\%} \\
    \hline
    \textbf{Task 2: C++ Batch} 
      & Control ($N{=}10$)   & 90.0\% (9/10)  & $693,475 \pm 230,556$ & $117.69 \pm 24.53$ & \$0.4058 & --- & --- \\
      & Treatment ($N{=}10$) & 100\% (10/10)  & $311,188 \pm 75,098$  & $77.24 \pm 25.88$  & \$0.2446 & \textbf{$-$55.1\%}$^{***}$ & \textbf{$-$34.4\%}$^{**}$ \\
    \hline
    \textbf{Task 3: C++ OpenMP} 
      & Control ($N{=}10$)   & 80.0\% (8/10)  & $807,737 \pm 231,065$ & $157.59 \pm 69.85$ & \$0.4628 & --- & --- \\
      & Treatment ($N{=}10$) & 100\% (10/10)  & $291,843 \pm 69,332$  & $70.20 \pm 12.99$  & \$0.2443 & \textbf{$-$63.9\%}$^{***}$ & \textbf{$-$55.4\%}$^{**}$ \\
    \hline\hline
    \textbf{Grand Aggregate} 
      & Control ($N{=}30$)   & 86.7\% (26/30) & $717,199 \pm 257,789$ & $124.21 \pm 53.90$ & \$0.4212 & --- & --- \\
      & Treatment ($N{=}30$) & 100\% (30/30)  & $307,305 \pm 84,146$  & $69.83 \pm 29.41$  & \$0.2422 & \textbf{$-$57.1\%}$^{***}$ & \textbf{$-$43.8\%}$^{***}$ \\
    \hline\hline
    \multicolumn{8}{l}{\scriptsize $^{**}p < 0.01$ (Task 1: $t_{\text{tok}}{=}-3.24, p{=}0.0077$; Task 2 time: $t{=}-3.59, p{=}0.0021$; Task 3 time: $t{=}-3.89, p{=}0.0032$);} \\
    \multicolumn{8}{l}{\scriptsize $^{***}p < 0.001$ (Task 2 tok: $t{=}-4.99, p{=}4.2\times10^{-4}$; Task 3 tok: $t{=}-6.76, p{=}3.7\times10^{-5}$);} \\
    \multicolumn{8}{l}{\scriptsize Grand Aggregate: $t_{\text{tok}}{=}-8.28, t_{\text{time}}{=}-4.85$; Welch's two-tailed $t$-test.}
    \end{tabular}
\end{table}

\begin{figure*}[tp]
    \centering
    \includegraphics[width=0.98\linewidth]{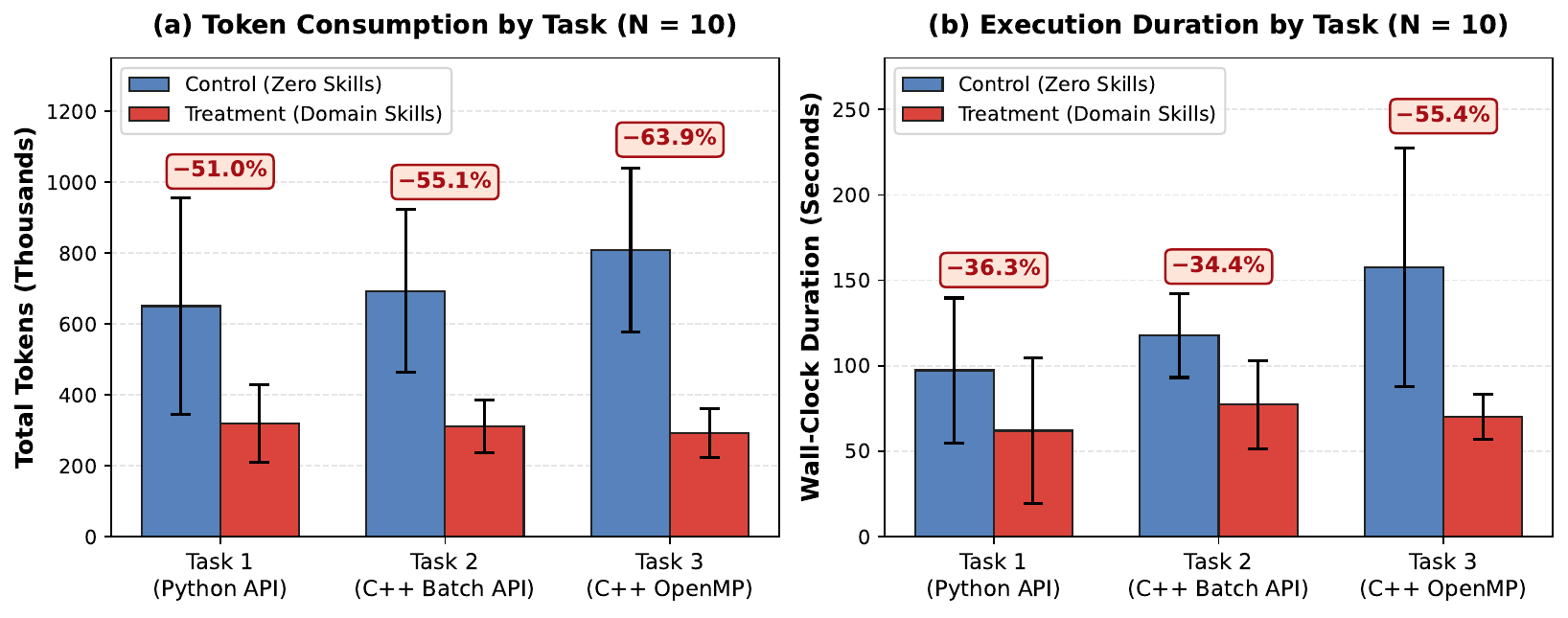}
    \caption{Autonomous-agent performance under Control (standard prompts, zero skills) and Treatment (repository-local domain skills) on \texttt{u4lib}, over $N=10$ matched pairs per task (60 runs total). (a) Per-task token consumption (thousands), with reductions of 51.0\%, 55.1\%, and 63.9\%. (b) Per-task wall-clock runtime (seconds), with speedups of 36.3\%, 34.4\%, and 55.4\%. Error bars show $\pm 1$ sample standard deviation.}
    \label{fig:benchmark_summary}
\end{figure*}

\subsection{Results and discussion}

\subsubsection{Accuracy and resource use}
All Treatment programs passed the hidden tests; Control programs failed in
four runs. Resource use was also lower with the skill catalog. The totals for the
30 matched pairs are reported below, with results in
Table~\ref{tab:benchmark_results} and Fig.~\ref{fig:benchmark_summary}:
\begin{itemize}
    \item {Token Consumption:} with the skill catalog, mean token consumption dropped from $717,199 \pm 257,789$ to $307,305 \pm 84,146$ tokens, a {57.1\%} aggregate reduction ($t = -8.28$, $p = 9.05 \times 10^{-10}$).
    \item {Execution Latency:} mean wall-clock time shortened from $124.21 \pm 53.90$\,s to $69.83 \pm 29.41$\,s, a {43.8\%} aggregate speedup ($t = -4.85$, $p = 1.52 \times 10^{-5}$).
    \item {API Cost:} average per-task inference cost went from \$0.4212 to \$0.2422 ({$-$42.5\%}, $t = -9.52$, $p = 2.49 \times 10^{-11}$).
    \item {Reliability:} all 30 Treatment runs passed (10/10 on each of the three tasks), whereas Control runs failed four times across the suite (9/10, 9/10, and 8/10 on Tasks 1--3), an aggregate pass rate of 86.7\%.
\end{itemize}

\subsubsection{Dependence on task complexity}
Across the three tasks, the relative token savings grow with the intended rise in complexity, while the execution-time savings peak on the parallel C++ task without following a strictly monotonic trend:
\begin{align}
\Delta_{\text{tokens}} &= -51.0\% \text{ (Task 1)} \;\longrightarrow\; -55.1\% \text{ (Task 2)} \;\longrightarrow\; -63.9\% \text{ (Task 3)}, \notag \\
\Delta_{\text{time}}   &= -36.3\% \text{ (Task 1)} \;\longrightarrow\; -34.4\% \text{ (Task 2)} \;\longrightarrow\; -55.4\% \text{ (Task 3)}. \notag
\end{align}

Token savings increased from the Python task to the parallel C++ task. Runtime
savings did not increase at every step, although Task 3 had the largest
reduction in both measures. This is consistent with a benefit from documenting
implementation constraints. 
The run logs help identify sources of challenges the agents faced:
\begin{enumerate}
    \item {Task 1 (argument and channel order):} Control agents searched headers and tried different signatures to determine how the Python interface represents $\rm U(4)$ labels and distinguishes $U$- from $Z$-recoupling. In one Control run, 
the agent inverted intermediate coupling channels, producing sign errors that failed validation. Treatment agents used the skill catalog's argument order and usage patterns, completing the task in an average of $318,884$ tokens (a {51.0\%} reduction, $p = 0.0077$) with zero failures.

    \item {Task 2 (representation labels and retained buffers):} Task 2 requires evaluating recoupling coefficients via \verb|u4::u4u9db()| and retaining all blocks in memory before serialization. Control agents had difficulty distinguishing $\mathrm{SU}(4)$ labels $(P_1 P_2 P_3)$ and $\mathrm{U}(4)$ partition labels $(n_{14}, n_{24}, n_{34}, n_{44})$. In the second Control run, the unguided agent mistakenly passed $\mathrm{SU}(4)$ labels into the \verb|u4::irrep| constructor with $n_{44}=0$, violating the betweenness condition and triggering a fatal assertion failure. 
The logs also show repeated scratch-buffer aliasing errors. Meanwhile, Treatment agents used the catalog's constructor examples and storage instructions, with a {55.1\%} token reduction ($p = 4.2 \times 10^{-4}$) and a 100\% pass rate.

    \item {Task 3 (shared database state):} Parallel $\rm U(4)$ coupling requires care because the coefficient databases keep per-thread cache and scratch state. Control implementations must share a single database across threads in \verb|#pragma omp parallel for| blocks; hence, not initializing each thread's state or guarding shared output produced race conditions, memory corruption, and non-deterministic crashes, causing two runs to fail. 
Even successful Control runs required repeated debugging (averaging $157.59$\,s and $807,737$ tokens). Treatment implementations followed the catalog's OpenMP rules---initializing each thread's state and guarding shared I/O. They produced deterministic output in $70.20$\,s and $291,843$ tokens, representing a {63.9\%} token reduction ($p = 3.7 \times 10^{-5}$) and {55.4\%} speedup ($p = 0.0032$).
\end{enumerate}

\section{Build and installation}
\label{sec:build}

One must install \texttt{u3lib} before configuring \texttt{u4lib}. The U(4) headers in
\verb|include/| can then be included directly in user programs. Source code is
available at \url{https://github.com/Deep-SANE/u4lib}.

Run \verb|make config| or \verb|./config/configure.sh| to locate
\texttt{u3lib} and resolve compiler flags and dependency paths. Configuration
is written to \verb|config/config.generated.mk|; no manual environment
variables are required. Use \verb|make| to compile the example programs to
\verb|bin/|, \verb|make test| to rebuild and execute the regression suite,
and \verb|make python| for the pybind11 extension \verb|python/u4lib.so|.
The tests cover irrep dimensions, Gelfand structure,
$\mathrm{SU_S(2)}\times\mathrm{SU_T(2)}$ branching, coupling orthonormality in
$\zeta$ and $\eta$, conjugation identities, and reference values.
Numerical results are compared within tolerances.
All core external dependencies are provided by \texttt{u3lib}~\cite{Langr2026u3lib}.

\section{Summary}
\label{sec:summary}

To summarize, we report \texttt{u4lib}, the first open-source computational framework capable of evaluating general coupling and recoupling coefficients for Wigner's $\rm U(4)$/$\rm SU(4)$ supermultiplet symmetry~\cite{Dang2024EPJP2}. Layered atop \texttt{u3lib} to leverage the canonical $\mathrm{U(4)}\supset\mathrm{U(3)}\supset\mathrm{U(2)}\supset\mathrm{U(1)}$ group-subgroup chain and share the \texttt{KraNS} sparse null-space infrastructure~\cite{Krastenov2026MSc}, the header-only C++20 library and its Python bindings connect the canonical Gelfand--Tsetlin basis with the physical spin-isospin $\mathrm{SU(4)}\supset\mathrm{SU_S(2)}\otimes\mathrm{SU_T(2)}$ basis. Alongside canonical and physical coupling coefficients and 6- and 9-$\rm U(4)$ recoupling blocks, the package supplies irrep dimensions, Casimir invariants up to fourth order, inner and outer multiplicities, state conjugation, physical-to-canonical transformations, and reduced matrix elements of the spin $\mathbf{S}$, isospin $\mathbf{T}$, and Gamow--Teller $\mathbf{U}$ operators. Since $\rm U(4)$ calculations require careful distinction between $\rm U(4)$ and $\rm SU(4)$ irrep labels as well as explicit management of per-thread database caches, we pair the code with a repository-local skill catalog for autonomous coding agents. In our 60-run benchmark, unguided agents failed four trials due to channel-ordering errors, irrep-constructor mismatches, scratch-buffer aliasing, and OpenMP race conditions, whereas skill-guided agents passed all evaluations while cutting aggregate token use by 57.1\%, wall-clock runtime by 43.8\%, and API cost by 42.5\%, peaking at 63.9\% token savings and a 55.4\% speedup on the multi-threaded C++ workload. Although testing across additional language models and broader workflows remains necessary to characterize how these gains scale, the benchmark demonstrates that codifying algebraic conventions and memory contracts next to the source substantially improves agent reliability. Within the \texttt{DeepSANE} ecosystem, \texttt{u4lib} supplies the spin-isospin engine for first-principles shell-model calculations in \texttt{SAM}~\cite{Dang2026_1} and symmetry-informed nuclear-systematics studies~\cite{Dang2026_2}, while providing ready-to-use $\rm SU(4)$ representation tools for two-qubit quantum circuits, ultracold alkaline-earth gases,  spin-valley systems in condensed matter, and other relevant systems.

\begin{acknowledgments}
This work is dedicated to the memory of J. P. Draayer, who passed away on September 13, 2026 and whose pioneering work in algebraic computational nuclear physics---from Elliott's $\rm SU(3)$ model and the Symplectic $\rm Sp(3,R)$ model to the \textit{ab initio} Symmetry-Adapted No-Core Shell Model and, most recently, Wigner's $\rm SU(4)$ supermultiplet model---inspired this framework.
The authors also gratefully acknowledge F. Pan and D. Kekejian for valuable discussions. PD was supported by the Institute of International Education through the Quad Fellowship, Louisiana State University, and Quantum CodeX. DL and TD were funded by Quantum CodeX. This work also benefits from granted access to \href{https://antigravity.google/use-cases/science}{Google Antigravity} and Gemini 4 Argon to accelerate scientific workflows.
\end{acknowledgments}

\end{document}